\documentclass{phyeauth}
\usepackage{graphicx}
\usepackage{amsmath}
\usepackage{amssymb}

\newcommand{\bea}{\begin{eqnarray}}
\newcommand{\eea}{\end{eqnarray}}

\begin{document}

\begin{frontmatter}

\title{Stationary-state electronic distribution in quantum dots}

\author[fzu]{Karel Kr\'al \thanksref{thank1},\thanksref{thank2}}
\author[fzu]{Petr Zden\v ek}

\address[fzu]{Institute of Physics, Academy of Sciences of Czech Republic, 18221 Prague 8,
Na Slovance 2, Czech Republic}

\thanks[thank1]{
Corresponding author. E-mail: kral@fzu.cz}
\thanks[thank2]{Support of K. K. from project AVOZ1-010-914 is acknowledged.}

\begin{abstract}
We wish to draw an attention to a non-gibbsian behavior of
zero-dimensional semiconductor nanostructures, which appears to be
manifested in experiments by an effect of incomplete depopulation from
electronic excited states or by an effect of up-conversion of
electronic level occupation after preparing the system in the ground
state of electronic excitation. In the present work the effect is
interpreted with help of electron-LO-phonon interaction, which is
supposed to play a role in these structures in the form of
multiple-scattering of electron on the optical phonons. Quantum kinetic
equation describing the process of electronic ralaxation with the
inclusion of electronic multiple scattering on phonons is considered.
The multiple electron scattering interpretation of the effect is
supported by pointing out a considerable degree of agreement between
the theoretical picture presented and a rather extensive amount of
existing experimental data.
\end{abstract}

\end{frontmatter}


\section{Introduction}
\label{sec:Introduction}

From the point of view of  mean free path of charge carriers the
semiconductor nanostructures are small objects. Thanks to this size of
nanostructures  charge carriers can repeat their elastic scattering at
the nanostructure boundaries many times and return back to a scatterer
inside the nanostructure, to perform in this way coherent
multiple-scattering acts. Such multiple-scattering processes are not
usually regarded as important in bulk semiconductor samples. In the
latter, a charge carrier makes a single scattering act and leaves the
target to infinity. When the scattering target in a nanostructure is
the phonon system, like the longitudinal-optical (LO) phonons in the
lattice of, say Gallium Arsenide (GaAs), the multiple scattering
process can lead to a creation of multi-phonon states in the system of
lattice vibrations, analogical to multi-photon states in the light of
laser. Making an analogy between the multi-phonon states and the
multi-photon states in the laser light, we may expect the multi-phonon
states to manifest themselves as macroscopic oscillations of the whole
atomic lattice of the nanostructure, making the environment, in which a
charge-carrier moves, to be a system with non-conservative property
from energy point of view. From such reasons we may expect unusual
properties of semiconductor nanostructures, as compared to ordinary
bulk sample properties.

This feature of electronic scattering on phonons in quantum dots has
been recently incorporated in the description of electron relaxation
theory \cite{my1998,Tsuchiya}. This theory has been applied to describe
the characteristics of electronic energy relaxation from higher energy
states of electronic excitation in quantum dots
\cite{my1998,Tsuchiya}. Rather recently, the theory has been
applied to a description of experimental effects of incomplete
electronic relaxation and up-conversion of electronic level occupation
in quantum dots \cite{ieee2004,ss2004}. A brief comparison of the
theoretical data with experiments will be presented in this work
emphasizing in this way a support which the experiments appear to
provide to the multiple-scattering theoretical interpretation,
postponing at this stage a more detailed analysis of the kinetic
equation properties to a later work.

In addition to comparing the theoretical transport properties with
experiment we perform a comparison of electronic spectral density with
experiment and direct numerical computation results.  In particular,
measurements of optical line-shape of luminescence from exciton ground
state in individual quantum dots allow us to compare with the
theoretical optical line-shape properties. Besides such a comparison
with experimental data, we also remark on comparison with recent
numerical computation published recently in paper \cite{Vasilevskiy},
in which electronic spectral density was computed in quantum dots
numerically from definition of electronic spectral density function.

In order to outline the theory under consideration in a necessary
detail, the model of electronic excitation in a quantum dot will be
specified and the kinetic equation describing the relaxation process
will be reminded, together with specifying the time scales which are
available for the description of effects under study. Properties of the
theoretical model of electronic relaxation, namely the properties of
limiting stationary state distribution of electron in a quantum dot,
will be characterized numerically.

In the summary we specify the implications which the theory, together
with the comparison with experiments, appears to suggest with regard to
statistical properties of nanostructures.

\section{Theoretical model of electronic relaxation in quantum dots}
\label{sec:Model}

Experiments on electronic relaxation in semiconductor quantum dots show
that this process usually occurs at the time scale of picoseconds
\cite{Nozik}. If quantum dot were not in contact with electromagnetic
field, the electron-phonon system of quantum dot would achieve a
stationary state within a period of picoseconds. If the state of the
phonon system were kept steady, we would expect that the electronic
subsystem statistical distribution equilibrates with the phonon system.
In this situation we could then use the word 'stationary' for the word
'equilibrium' and vice versa. Thanks to the coupling of electronic
excitations in quantum dots to electromagnetic field the above
mentioned stationary state slowly decays with a lifetime, which is
usually of the order of nanoseconds. With respect to the picosecond
relaxation time we shall regard the time period of nanoseconds as long
for achieving the stationary state due to electron-phonon interaction.
At the time scale of picoseconds we shall therefore not take into
account the interaction with electromagnetic field explicitly.

In a semiconductor quantum dot, or in a nanoparticle, electronic
excitation may have eigenstates with energy below the  outside
electronic potential of this structure. In quantum dot samples the
bound states may be controlled by choosing a suitable combination of
two materials of such a heterostructure \cite{Yoffe}, with appropriate
combination of energies of valence and conduction bands of hole and
electron states. Such a heterostructure may be e. g. a GaAs qauntum dot
surrounded by the material of AlGaAs
\cite{Nozik,Yoffe}. Let us simply speak only about quantum dots in what
follows, although much of our arguments may be related also to other
types of nanostructures. As for the electronic states in quantum dots,
we usually deal with the most simple electronic excitations, in which a
single electron is present in conduction band bound states of quantum
dot, while the valence band states are occupied by a single hole
particle.

The electron and the hole mutually interact by electrostatic forces and
both of them are  coupled to lattice vibrations. The well-known
differences between these two particles allows us to make a simplifying
assumption about the influence of holes. Namely, it is well-known that
the holes relax in a rather fast way to their ground state and they are
much heavier than the electrons. Assuming that the holes are heavy
enough we completely neglect their motion, assuming that they simply
contribute to the effective potential in which the electron moves. In
this way the electronic excitation as a whole can be represented simply
by the electron in the conduction band states of a quantum dot. For the
sake of simplicity we then speak here about electron instead of
speaking about electron-hole complex, or about an exciton.

The three-dimensional potential well, to which the electronic motion is
confined when moving in the bound states, can be simply represented by
the infinitely deep cubic potential well, with the lateral size $d$, in
which the electron has the effective mass of the bottom of the
conduction band $\Gamma$-valley of GaAs. We shall confine ourselves to
the two lowest-energy electron bound states in this potential, namely
to the ground state, with unperturbed energy $E_0$, taking $E_0=0$ in
this work, and one of the triple-degenerate lowest-energy excited
states, with energy $E_1$.

The single electron moving inside the quantum dot is assumed to
interact with bulk modes of longitudinal optical (LO) phonons of the
lattice. The electron-phonon coupling is provided by the well-known
Fr\"{o}hlich's coupling \cite{Callaway}. The corresponding material
constants will be those of GaAs crystalline material. This
electron-phonon coupling is usually found to be the strongest coupling
of charge carriers in polar semiconductors
\cite{JacoboniReggiani} to lattice modes.
The Hamiltonian of the system can be found in the reference
\cite{ss2004} and will be not repeated here.
Throughout this work we assume that the system of LO
phonons is kept at equilibrium temperature $T_{LO}$.

We shall specify the state of the electronic system by electron state
occupations $N_i$, $i=0,1$, $N_i$ being mean value of electronic number
operator  of electrons in state $i$, $N_0+N_1=1$. $N_0$ is therefore
the occupation of the lower energy electronic state with energy $E_0$.
We use the method of nonequilibrium Green's functions and apply the
simple Kadanoff-Baym ansatz
\cite{KadanoffBaym}  being aware of its certain shortcomings from
the point of view of causality violation \cite{HaugKoch}.
 We  include the electron-LO-phonon interaction into
 the electronic self-energy in the form of  the
self-consistent Born approximation \cite{PRB1998}. Then we get the
following quantum kinetic equation for time evolution of electronic
distribution due to electron-LO-phonon interaction :
\begin{eqnarray}
\frac{dN_1}{dt}=-\frac{2\pi}{\hbar}  \alpha_{01}
 \bigg[ N_1(1-N_0)  \nonumber \qquad \qquad \qquad \qquad \\
\times \Big( (1+\nu_{LO})  \nonumber
\int^{\infty}_{-\infty}dE\,\sigma_1(E)\sigma_0(E-E_{LO})
\label{rate} \nonumber \\
+\nu_{LO}\int^{\infty}_{-\infty}dE\sigma_1(E)\sigma_0(E+E_{LO})
\Big) \nonumber\\
- N_0(1-N_1)\nonumber \qquad \qquad \qquad \qquad \\ \times
\Big((1+\nu_{LO})
\int^{\infty}_{-\infty}dE\sigma_0(E)\sigma_1(E-E_{LO}) \nonumber \\
+\nu_{LO}\int^{\infty}_{-\infty}dE \sigma_0(E)
 \sigma_1(E+E_{LO})\Big) \bigg] .
\end{eqnarray}
In this equation $\nu_{LO}$ is Bose-Einstein distribution function of
LO phonons kept at temperature $T_{LO}$. The optical phonons are
assumed dispersionless. Instead of energy-conservation delta-function,
which would be present on the right-hand side of this equation in case
of approximating the self-energy by bare Born approximation, the
collision integral depends here on convolution of two electronic
spectral densities $\sigma_i(E)$ corresponding to states $i=0,1$. This
off-shell property of the collision integral occurs because of the
presently used self-consistent Born approximation, which, as said
above, can be seen as introducing effectively virtual multi-phonon
states together with a certain effective energy non-conservativeness of
the system in the present system of electron-phonon dynamics. Because
of the significant property of quantum dots, due to which electronic
motion is severely restricted in space in all three dimensions, the
self-consistent Born approximation, together with the off-shell
property of the collision integral, may appear to be substantial for
describing processes in quantum dots.

The electronic spectral densities are determined with help of
electronic Green's functions, which are again given by the electronic
self-energy. The self-consistent Born approximation to the electronic
self-energy is given by the following self-consistent equation for
retarded self-energy $M^R_n(E)$, in which $G^R_m(E)$ is retarded
electronic Greens's function in state $m$:
\begin{eqnarray}
M_n^R(E)\qquad \qquad \qquad \qquad \qquad \qquad \qquad \qquad \qquad
\qquad \nonumber \\
\qquad =\sum_{m}\alpha_{nm}
\big[ (1-N_m+\nu_{LO})G^R_m(E-E_{LO}) \qquad
 \nonumber \\
+(N_m+\nu_{LO})G^R_m(E+E_{LO})
 \big]. \qquad \qquad \qquad
 \label{SCB}
\end{eqnarray}
Despite the fact that the unperturbed Hamiltonian does not contain any
continua in the spectrum of its eigenstates, the solution of
(\ref{SCB}) taken to the limit of infinite number of iterations of the
 equation gives electronic spectral densities which are not completely
discrete. This property was demonstrated earlier
\cite{PRB1998}. In an intuitive way, this property of the solution can be
understood on the basis of multiple scattering motion of electron in a
quantum dot and a presence of virtual multi-phonon states. Parameters
$\alpha_{nm}$ determine the electron-phonon coupling. The reader is
referred e. g. to paper \cite{ss2004} and to references cited in.

\begin{figure}[tb]
  \begin{center}
    \includegraphics[angle=0,width=.45\textwidth]{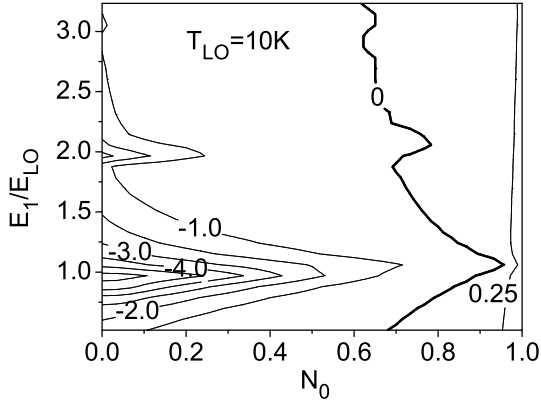}
    \caption{Contour plot of relaxation rate $dN_1/dt$ in units of ps$^{-1}$.
    Note the curve with
    index 0 marking such detuning $E_1/E_{LO}$ and occupation $N_0$
    at which relaxation rate is zero at the given temperature of lattice.}
    \label{fig:ratemap}
  \end{center}
\end{figure}

Earlier numerical evaluations of the relaxation rate $dN_1/dt$
(\ref{rate}) show that the electron relaxes quickly on picosecond time
scale without any strong dependence on the detuning $E_1/E_{LO}$
between the electron energy level separation and the energy of the
optical phonon \cite{my1998,Tsuchiya}. On the basis of standard
experience we might expect that when the electron is prepared in the
higher-energy state $n=1$, with energy $E_1$, the electron-phonon
interaction will at the end lead to the limiting occupation $N_0=1$, at
zero temperature of lattice. However, the above given relaxation rate
formula and, as will be shown in the next section, a number of recent
experiments too, do not seem to provide such a simple picture of
electron energy relaxation in quantum dots. We shall comment on this
question in what follows, adding also a brief review of experiments in
the next section.

Let us characterize first the solution of relaxation rate equation
(\ref{rate},\ref{SCB}) for some special situations. Let us consider the
limiting state at which electronic distribution reaches stationary
distribution at a given temperature $T_{LO}$ of the lattice. This state
can be obtained by letting evolve in time the electron distribution
according to the differential equation (\ref{rate},\ref{SCB}), or
simply evaluating  the  relaxation rate, finding out, for given
conditions, when this quantity happens to be zero. In
Fig.~\ref{fig:ratemap} we present a contour plot of the relaxation rate
$dN_1/dt$ as it depends on the detuning $E_1/E_{LO}$ and occupation
$N_0$. The curve marked with index 0 displays  conditions under which
the relaxation rate is zero. Therefore, this curve gives the stationary
states to which the system evolves in the limit of long time. We see
that in some dependence on the detuning the stationary value of the
occupation $N_0$ of the lower energy state is about 0.7, with a certain
strong extremum near the detuning equal to 1.

\begin{figure}[tb]
  \begin{center}
    \includegraphics[angle=0,width=.45\textwidth]{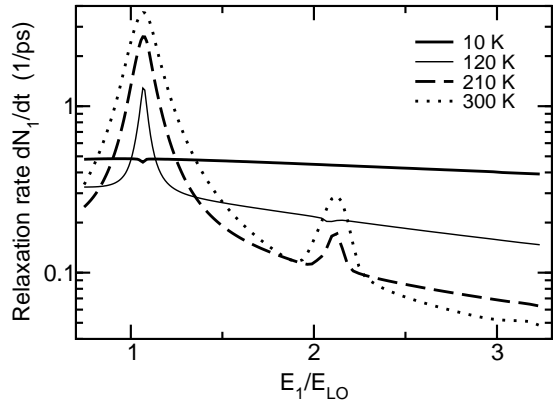}
    \caption{ Rate $dN_1/dt$ of up-conversion of electronic
    occupation from the fully occupied lower energy state ($N_0=1$)
    to the upper
    energy state.}
    \label{fig:rate-dole}
  \end{center}
\end{figure}

When the electron is prepared in the low energy state at the beginning
of evolution under electron-phonon interaction, the relaxation process
displays a tendency to increase the occupation of the upper energy
state. Detailed calculation of the evolution of the system shows
\cite{ieee2004} that this process occurs on the time scale of
picoseconds. The relaxation rate describing this process of
up-conversion in the state when $N_0=1$  is presented in
Fig.~\ref{fig:rate-dole}. In this figure the rate of up-conversion
displays resonance structure at  large temperatures of lattice, while
the rate is nearly structureless in the limit of zero temperature of
the lattice. The effect of nonzero rate of up-conversion near
$T_{LO}=0$ can be explained by the nature of the electron-phonon
interaction, which leads to multiple-phonon states in the quantum dot,
which leads to an non-conservativeness of the effective Hamiltonian in
which the electron moves. Technically, the term, which is responsible
in formula (\ref{rate}) for this zero temperature up-conversion, is
that proportional to the factor $N_0(1-N_1)(1+\nu_{LO})$. This term
corresponds to an emission of phonons in this process. We give the same
explanation of the plausibility of the simultaneous electronic
up-conversion and phonon emission: because of the multiple-scattering
and the off-shell nature of the process we do not control the
conservation of energy in formula (\ref{rate}). Because of the
off-shell property of the multiple-scattering electron-phonon
scattering processes, the comparison of unperturbed energies of
electrons and phonons may not give us here a sufficient control over
energy balance in the present model and approximation. Obviously, this
point deserves an attention in future work.

\begin{figure}[tb]
  \begin{center}
    \includegraphics[angle=0,width=.45\textwidth]{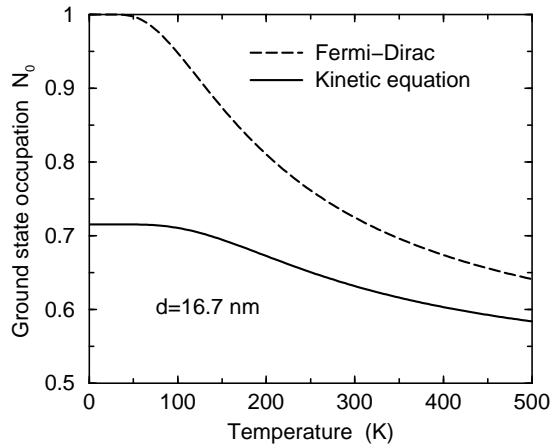}
    \caption{ Fermi-Dirac occupation $N_0$ of lower energy state
     in  quantum dot
    with  lateral size of $d$=16.7 nm (dashed). Full line gives
    occupation  $N_0$ calculated from the kinetic equation
    in  steady state.}
    \label{fig:FD}
  \end{center}
\end{figure}

In Fig. (\ref{fig:FD}) we show the limiting stationary occupation $N_0$
of the lower energy state calculated from kinetic equation
(\ref{rate},\ref{SCB}), as a function of the lattice temperature
$T_{LO}$. We see that the amount of electronic density staying in the
low energy state is much lower than the amount given by Fermi-Dirac
distribution at the same temperature $T_{LO}$ of lattice. This effect
may be interpreted as a non-gibbsian behavior of electron-phonon system
in semiconductor quantum dots. This effect is obtained within our
present simple theoretical model with the present simple approximations
to electronic self-energy. The reader is referred to papers
\cite{ieee2004,ss2004} for other numerical results on the
topic of time dependence of electronic relaxation in quantum dots.

The model and the kinetic equation  deserve themselves a further
theoretical analysis, which should test e. g. the stability of the
effect with respect to further extending the model of quantum dot
towards a sufficiently realistic model of quantum dot, and also with
respect to improvements of the approximations leading to the kinetic
equation. In particular, we would like to gain a better understanding
of the energy balance in processes under consideration. Instead of
proceeding further in this respect we give way in this work to
presenting a summary of the present state of agreement between the
theory and the available experimental data.

\section{Experiments on electronic relaxation}
\label{sec:Experiments}

The theoretical value of the relaxation rate, given by
(\ref{rate},\ref{SCB}) and being often in the range of 1/ps, agrees
well by the order of magnitude with a large number of experimental
measurements on the electron energy relaxation in quantum dots (see e.
g.
\cite{Nozik}). In addition to this, the relaxation rate depends to some
extent on the detuning $E_1/E_{LO}$, giving resonance spikes on the
relaxation rate (\ref{rate}) at the detuning approximately equal to an
integer
\cite{my1998}. The experimental data on photoluminescence excitation
(PLE)  display often a certain resonance structure with a periodicity
equal to optical phonon energy along the photon energy variable axis
\cite{LedentsovSemi33,LedentsovSemi32}. The presence of the
resonance structure in the electronic relaxation rate is in agreement
with the purely optical experimental method used sometimes for
detection of the presence of quantum dots in the grown samples (see
ref.~\cite{LedentsovSemi32}, section 3.3).  The experimental electron
relaxation rate and the  resonance structure can be related to recent
theoretical calculations of the electron energy relaxation in quantum
dots based development of polaron concept in quantum dots and on
considering finite lifetime of optical phonons (e. g.
\cite{Verzelen2002}). This latter approach, in which the question of
polaron on quantum dots has been developed considerably, seems to give
agreement with experiment in certain restricted intervals of the
magnitude of the detuning. The up-conversion effect has not been
reported in the polaron approach as yet.

The electronic spectral density itself, as it is given by the
self-consistent equation (\ref{SCB}), gives, at integer level
occupation, the spectral line shape having the  functional dependence
of the type $1/\sqrt(E)$ at low temperatures
\cite{PRB1998}, which has zero width at $T_{LO}=0$
in the sense of full width at half maximum. The temperature dependence
of the main feature of the spectral density
\cite{my-shape} is found in good agreement with experiments \cite{Ota1998}.
The electronic spectral density line shape profile of the lower energy
electronic state in our model, regarded to correspond to the optical
line shape of the ground state electronic excitation in the quantum
dot, has a shoulder on the lower energy side, with the shoulder width
of the order of milielectronvolts. Such a characteristic line-shape of
the lowest energy optical transition has also been recently detected in
single-dot experiments on polar semiconductor quantum dots
\cite{Pelucchi2003,Taylor2004} and on Si nanoparticles \cite{ValentaSi}.

Still another supporting argument is provided to the presently used
self-consistent Born approximation to the self-energy by the recently
published numerical analysis  based upon taking into account the
electron-LO-phonon coupling in numerical evaluation of the electronic
spectral density from definition \cite{Vasilevskiy}. The numerical
results of this study, although they cannot provide continuous spectral
density because the calculation is based on the phonon Fock's space
with only finite number of phonons included, avoiding any inclusion of
coherent states of the phonon modes, gives a structure of discrete
lines of the spectral density, the envelope of which nevertheless
appears to remind interestingly the results of the presently used
self-consistent Born approximation. This remarkable agreement of the
spectral density with experimental data, and even with results of
direct numerical analysis, supports the plausibility of choosing
self-consistent Born approximation to include the electron-LO-phonon
coupling in the present transport equation.

Let us turn our attention now to experiments related to the electronic
transport measurements of effects like the electronic up-conversion  in
quantum dots.

There is a relatively large number of experiments on anti-Stokes
emission and electronic up-conversion in nanostructures and other
systems in scientific literature. Some of them were mentioned in papers
\cite{ieee2004,ss2004} and some will be summarized elsewhere. Here we
pay attention to the following ones. In the experimental papers which
we want to mention the measurement of luminescence provides data, which
can be related to occupation of bound states of the electronic system
in  quantum dots.

When we have in mind excited state of an exciton in a quantum dot, we
shall speak about excited  state. We shall speak about ground state,
when having in mind the lowest energy state of electronic excitation in
the dot (an exciton). One of the most interesting papers on incomplete
electronic level depopulation is the measurement of time resolved
luminescence signal \cite{Urayama,Urayama2}, measured at 6 or at 40 K
on InGaAs quantum dots, where depopulation of an excited state of
electronic excitation in quantum dots is determined. It is shown that
within picoseconds the excited state is being depopulated, but the
depopulation process stops, leaving the depopulation of excited state
incomplete. The amount of excited state population at which the
depopulation stops is in the order of tens of percents. Similarly such
process of incomplete depopulation of excited state was detected in
reference
\cite{Quochi}, in which the authors
 measured the same process at 290 K in InAs/GaAs quantum dots.
After exciting the exciton ground state, they report appearance of
emission from excited state of exciton, which is detected about 60 ps
after excitation of the ground state. The authors regard the process of
appearance of the excited state luminescence as unusually fast for to
be caused by simply a thermal population of the exciton excited state.
According to their measurement, the luminescence from the resonantly
excited ground state decreases quickly on a picosecond scale, with the
ground state signal becoming weaker by several tens of percents, after
which the excitation decays on the scale of nanoseconds. Let us
emphasize that the remarkable fast process of partial depopulation of
the resonantly excited exciton ground state can be explained by the
up-conversion process provided by the present theory.

Similarly an incomplete depopulation has been detected in CdSe quantum
dots at 300 K \cite{Guyot-Sionnest}. The authors also report their
observation about a certain  difficulty to achieve in their experiment
a state with an integer occupation of quantum dot levels. The integer
level population is unstable even within the present theory.

Diener {\it et al}. measured anti-Stokes luminescence at 4 K in coupled
Si nanocrystals \cite{Diener}. The work puts emphasis on the excitation
intensity dependence of the up-converted emission signal. Although the
measured effect is interpreted in terms of Auger mechanism of
up-conversion of electronic system to excited states, the data
presented do not seem to display any strong excitation intensity
dependence, leaving space for an alternative interpretation too.
Similarly, measurements of emission from excited states was reported in
the reference \cite{Ikeda}. The authors also have put emphasis on
excitation intensity dependence of excited state luminescence spectrum.
Although a factor of 90 in variation of excitation intensity did not
seem to influence the spectra significantly enough, the experiment was
interpreted in terms of Auger mechanism of luminescence up-conversion.
In a quite similar way, excitation intensity dependence of the excited
states luminescence was measured some time ago without showing a strong
appearance of such a dependence in experiment on GaAs/AlGaAs quantum
dots
\cite{Bockelmann}. Also this measurement was interpreted in terms of
Auger mechanism of electronic state occupation up-conversion. In
experimental paper reported in reference
\cite{Zhang} the effect of up-conversion is nicely demonstrated in the
low excitation intensity limit in InGaAs/GaAs quantum dot systems.

Even the effect of the up-conversion to wetting-layer states has been
detected, using photoluminescence and photoluminescence excitation
measurement
\cite{Kammerer}. Let us remark at this point that recently we have
given theoretical arguments in favor of existence of such an effect in
dense quantum dot samples excited to a sufficiently large degree of
electronic excitation \cite{ieee2004}.

Let us also remark that at the low level of electronic excitation in
luminescence measurements the present multiple-LO-phonon scattering
mechanism may be at least a suitable alternative to the Auger mechanism
of the up-conversion, while at larger excitation intensities, under
which the Auger mechanism becomes possible at all, the phonon mechanism
may be at least contributing to the up-conversion mechanism. A more
detailed relation between the role of the two mechanisms of
up-conversion has not yet been shown.

\section{Summary}
\label{sec:Summary}

In the experiments which we mentioned above the electronic subsystem
displays a tendency to reach a state with a non-integer occupation of
electronic states, even in the limit of zero temperature of lattice.
The numerical results presented here display this tendency as well. The
experiments also show that the up-conversion effect can be observed
even at low intensity of electronic excitation of quantum dot sample,
which restricts somewhat the interpretative potential of Auger
mechanism in the present context. Besides the agreement of the
magnitude of electron energy relaxation rate with experiment, also the
presence of resonant features of the relaxation rate dependence on
level detuning appears in correlation with the earlier reported
practical method of detection quantum dots in a sample. Also the
computed line shape and shoulder width compares well with experimental
and other numerical data. Finally, the luminescence experiments on
incomplete depopulation and up-conversion appear to agree well with the
trends indicated in multiple-LO-phonon scattering mechanism numerical
results in quantum dots.

The above presented relation between the theoretical and the
experimental data provides supporting arguments in favor of the
presently discussed theoretical interpretation of the effects.
Undoubtedly, further experiments are needed in order to delineate more
clearly the validity of the present model and approximations to the
kinetic equation for electronic relaxation in quantum dots, including
the energy balance question. Nevertheless, providing that the mechanism
presented here will show up as a valid mechanism of electron-phonon
motion in quantum dots, these systems might represent a candidate for
the practical realization of a non-gibbsian behavior of a small system
in steady state. At the same time, even without speaking about a need
to understand nanostructures for the purpose of applications, this
small quantum mechanical system might be an interesting object for
further study of nonequilibrium processes and statistical physics in
small interacting systems.

In particular, the present relation between theory and experiment leads
to the conclusion that the steady state electronic distribution, which
may be achieved within picoseconds in nanostructures, is not given by
Gibbs distribution with the temperature equal to the temperature of the
lattice, but it should instead be determined by a solution of a kinetic
equation. In the present case of two-level electronic system such a
solution would then mean that the electronic temperature would not be
necessarily equal to that of the lattice modes with which the electrons
interact in this nanostructure. Beyond such a behavior of the system
there would then be a smallness of the system with respect to the
dephasing mean free path of electron, which conditions would make
situation open for an off-shell behavior of multiple scattering of
electrons. Because of some degree of carrier confinement present also
in other nanostructures, certain effects of a similar nature should not
be unexpected in them.

The volume of agreement between the present theoretical scheme and
experimental experience is rather broad, so that it may seem that a
mere accidental coincidence between theory and experiment is not
likely. At the same time questions may be arisen both about the theory
and the experiment. The behavior of charge carriers in transport
processes in quantum dots therefore deserves a further attention.



\end{document}